\documentclass[12pt]{iopart}
\usepackage{graphicx}
\usepackage{dcolumn}
\usepackage{bm}
\usepackage[usenames]{color}
\usepackage{ae, aecompl, aeguill}
\usepackage{graphics}

\newcommand{\2}{$_2$}
\newcommand{\RCo}{$R$Co$_2$}

\newcommand{\Tc}{$T_\mathrm{c}$}

\newcommand{\Tf}{$T_\mathrm{f}$}

\newcommand{\mCo}{$m_{\mathrm{Co}}$}

\newcommand{\muB}{$\mu_B$}

\newcommand{\XT}{$\boldsymbol{\chi}_{ac}(T)$}
\newcommand{\XReal}{$\chi_{ac}'$}
\newcommand{\XImg}{$\chi_{ac}''$}
\newcommand{\LL}{L$_{2, 3}$}

\newcommand{\Xw}{$\boldsymbol{\chi}_{ac}(\omega)$}
\newcommand{\w}{$\omega$}
\newcommand{\wt}{$\omega$$\tau$}
\newcommand{\GlP}{Griffiths-like phase}
\newcommand{\G}{Griffiths}
\newcommand{\CW}{Curie-Weiss}
\newcommand{\D}{Debye}
\newcommand{\DC}{Davidson-Cole}
\newcommand{\VF}{Vogel-Fulcher}

\begin{document}

\title{Observation of a \GlP\ in the paramagnetic regime of ErCo\2}

\author{Julia Herrero-Albillos,$^1$ Luis Miguel Garc\'{\i}a$^2$ and Fernando Bartolom\'{e}$^2$}
\address{$^1$ Department of Materials Science and Metallurgy, University of Cambridge, Pembroke Street, Cambridge, CB2 3QZ, United Kingdom} 
\address{$^2$ Instituto de Ciencia de Materiales de Arag\'{o}n, CSIC - Universidad de Zaragoza, Departamento de F\'{i}sica de la Materia Condensada, Pedro Cerbuna 12, 50009 Zaragoza, Spain}

\eads{\mailto{Julia.Herrero@unizar.es}, \mailto{luism@unizar.es} and \mailto{bartolom@unizar.es}}

\begin{abstract}

A systematic x-ray magnetic circular dichroism study of the paramagnetic phase of ErCo\2\ has recently allowed to identify the inversion of the net magnetization of the Co net moment with respect to the applied field well above the ferrimagnetic ordering temperature, \Tc. The study of small angle neutron scattering measurements has also shown the presence of short range order correlations in the same temperature region. This phenomenon, which we have denoted $parimagnetism$, may be related with the onset of a \GlP\ in paramagnetic ErCo\2. We have measured $ac$ susceptibility on ErCo\2\ as a function of temperature, applied field, and excitation frequency. Several characteristics shared by systems showing a \G\ phase are present in ErCo\2, namely the formation of ferromagnetic clusters in the disordered phase, the loss of analyticity of the magnetic susceptibility and its extreme sensitivity to an applied magnetic  field. The paramagnetic susceptibility allows to establish that the magnetic clusters are only formed by Co moments as well as the intrinsic nature of those Co moments.

\end{abstract}

\pacs{Valid PACS appear here}

\maketitle

\section{Introduction}

The study of \G\ phases \cite{Griffiths1969PRL23} in intermetallic compounds have reached much interest in the recent years. The extended \G\ model \cite{Bray1987PRL59, Bray1982JPC15} accounts for systems in which the magnetic correlations do not vanish completely at the first order phase transition, but are present at higher temperatures. The \GlP\ is then the region between the completely ordered state (i.e. above \Tc ) and the conventional disordered paramagnetic state. Experimentally a \GlP\ has been usually identified as the paramagnetic susceptibility shows deviations from a \CW\ law and extreme sensitivity to applied magnetic fields.
\cite{Salamon2003PRB68, Jiang2007PRB76, Ouyang2006PRB74, Li2007PRB75, Magen2006PRL96, Fan2007JAP101}
An increasing number of very interesting systems are being reported to show that kind of behavior. Among those, compounds that show giant magneto-caloric effect like Tb$_5$Si\2Ge\2,\cite{Magen2006PRL96} colossal magneto-resistance like La$_{1-x}$Sr$_x$MnO$_3$,\cite{Deisenhofer2005PRL95}  La$_{1-x}$Ca$_x$MnO$_3$,\cite{Salamon2002PRL88}  or La$_{0.7-x}$Dy$_x$Ca$_{0.3}$MnO$_3$,\cite{Yusuf2006PRB74} strongly correlated electron systems like CeNi$_{1-x}$Cu$_x$\cite{Marcano2007PRL98} or itinerant magnetic semiconductors like Fe$_{1-x}$Co$_x$S\2 \cite{Guo2008PRL100} are remarkable examples. In all those systems, the presence of a \GlP\ is related with the existence of short range order magnetic correlations, which is evidenced from the enhancement of the magnetic signal in small angle neutron scattering (SANS) measurements. It is very interesting therefore, to revisit those well studied materials in which there are experimental evidences of important short range correlations well above \Tc\ in order to determine whether or not they also show a \GlP\ behavior. What is more, a general question should be open: are \GlP s a common phenomena in a wide range of intermetallic compounds and what role they play in the magnetic properties of those compounds? In that sense, we present in this paper a study of the paramagnetic $ac$ susceptibility of the intermetallic compound ErCo\2\ in the search of the fingerprint of \GlP s.

Among the intermetallic compounds, the family of the \RCo\ are particulary interesting as they have a fairly simple structure, which facilitates the understanding or their magnetic properties. In fact they have been widely studied for decades as a model system of the magnetism of itinerant electron systems and metamagnetic processes.
However, it was not until recently that short range order correlations were shown to exist well within the paramagnetic phase of one of the members of the series. A SANS study in the compound ErCo\2\ showed the presence of magnetic clusters of around 8 \AA\ at temperatures as high as twice the ordering temperature when a magnetic field of 1 T is applied.\cite{Herrero-Albillos2007JMMM310}
Indeed, this behavior is related with the existence of the new magnetic phase reported in Ref.~ \cite{Herrero-Albillos2007PRB76}, where a new magnetic phase diagram for ErCo\2\ was proposed. We have denoted that new magnetic phase the $parimagnetic$ phase as -although no long range order exist in the compound- the Co moments are found to be, in average, oriented antiparallel to the Er moments.

The molecular field created by Er moments on the Co sublattice in ErCo\2\ is just above the critical value to induce the metamagnetic transition in the Co moments. Therefore, the Co sublattice is especially sensitive to variations of the external parameters. Below \Tc, the magnetic behavior is dominated by the Er sublattice, with an essentially temperature-independent magnetic moment of around 8.8 \muB\ per atom, while the Co sublattice is ordered antiparallel to the Er one through a first-order magneto-structural transition.\cite{Landolt-Bornstein, Moon-1965-A} In the paramagnetic phase, the nature of the cobalt magnetic moment is an especially interesting open question. A recent work by Liu and Altounian \cite{Liu-2006-A} predicted theoretically a transition from a low-spin state for Co ($\sim$~0.1~\muB) to the well known high-spin state of the ferrimagnetic phase ($\sim$~1~\muB) at the onset of the ordering temperature of ErCo\2. Moreover, whether or not an intrinsic magnetic moment in the Co sublattice exists has been a continuous matter of debate since the first susceptibility studies carried in the \RCo\ series. \cite{Gignoux-1976-A, Burzo-1972-B} The present study of the paramagnetic $ac$ susceptibility in ErCo\2\ will also deal with the existence and nature of Co moment in the paramagnetic phase.

The paper is organized as follows; we describe the synthesis and characterization of the samples in section \ref{sec:samplecharacterization}. The experimental results and discussion will be presented in section \ref{sec:Experimentalresults}: the $ac$ susceptibility measurements as a function of temperature and as a function of the excitation frequency will be shown in subsections \ref{subsec:XT} and \ref{subsec:Xw} respectively. The determination of the Co magnetic moment form the those data will be presented in subsection \ref{subsec:calcmoment}. Finally, in section \ref{sec:Conclusions}, we summarize the main results obtained.

\section{Sample Characterization}\label{sec:samplecharacterization}

\begin{figure}[!b]
\begin{center}
\includegraphics[width = 0.5\textwidth]{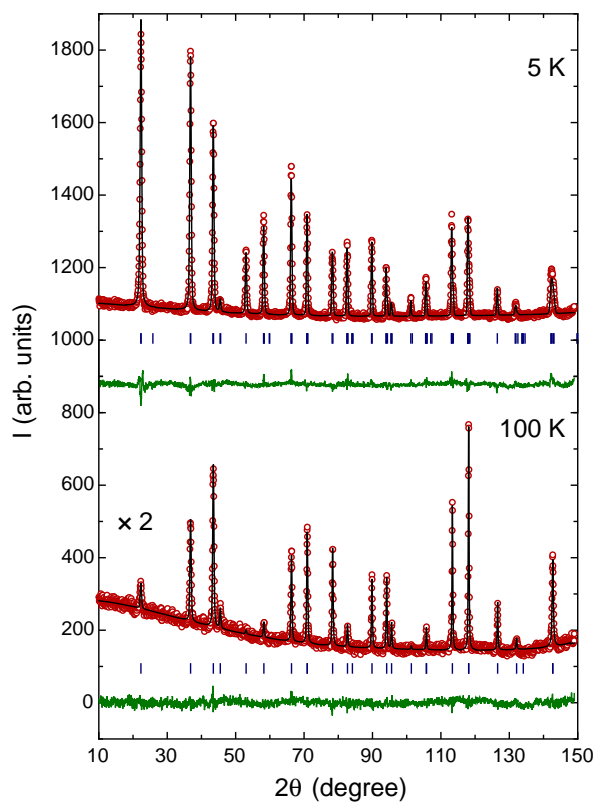}
\caption{(color online) ErCo\2\ neutron diffractograms at 5 K (upper panel) and 100 K (lower panel) at zero DC applied magnetic field. The empty circles and the continuous line are, respectively, the observed and calculated intensities from the Rietveld refinement. For the diffractogram at 5 K, the magnetic contribution has been included in the refinement. Vertical lines indicate the position of the Bragg peaks and the continuous lines below show the difference of the observed and calculated intensities. The diffracted intensity at 100 K has been multiplied by a factor of 2 for a better comparison.}
\label{1_DN}
\end{center}
\end{figure}

The ErCo\2\ samples are polycrystalline ingots and were prepared by melting the pure elements in an induction furnace under Ar atmosphere. The resulting ingots were further annealed under Ar atmosphere at 850 $^{\circ}$C for a week, wrapped in tantalum foil. Several ingots were synthesized in order to check the reproducibility of the phenomena studied. X-ray diffraction analysis at room temperature was performed on powdered samples to check their quality. Rietveld analysis of the diffractograms was perform using FullProf software\cite{Rodriguez-Carvajal-1990-A, Rodriguez-Carvajal-1993-A} assuring single phase samples with good crystallization and the expected cubic Fd$\overline{3}$m. No impurities were found within the 1\% accuracy of powder diffraction methods. The room temperature x-ray diffractogram together with the Rietveld refinement can be found in Ref.  \cite{Herrero-Albillos2006PRB73}. Neutron diffraction measurements were also performed in the sample at 100, 50 and 5 K. The neutron diffractogram for ErCo\2\ at 5 and 100 K are shown in Fig.~\ref{1_DN}. A low angle signal can be observed in the 100 K measurement, indicating the presence of short range correlations in the paramagnetic phase. The Rietveld refinement at 5 K confirms the ferrimagnetic coupling of Er and Co with magnetic moments for Er and Co of 8.84 $\pm 0.06$ \muB\ and 0.95 $\pm 0.03$ \muB\ respectively, and the space group R$\overline{3}$M.

\begin{figure}[!b]
\begin{center}
\includegraphics[width = 0.5\textwidth]{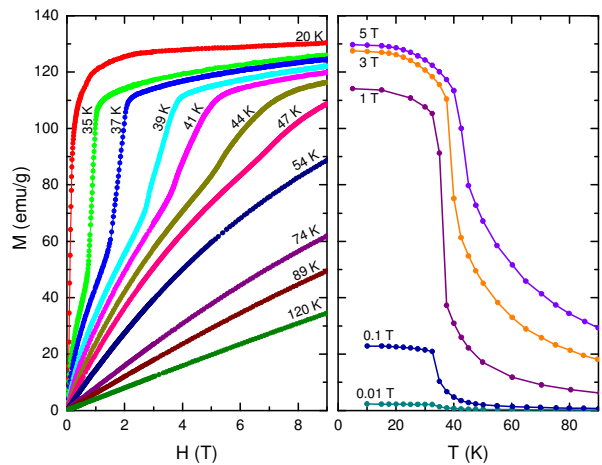}
\caption{(color online) Selected ErCo\2\ isothermal (left) and isofield (right) magnetization curves measured in SQUID and PPMS magnetometers.}
\label{2_MTHErCo2}
\end{center}
\end{figure}

A complete magnetic characterization has been performed on the ErCo\2\ samples. Magnetization measurements, $M(T,H)$, from 10 to 300 K and up to 5 T were performed in a SQUID Quantum Design magnetometer and up to 9 T in a commercial Quantum Design extraction magnetometer. Selected $M(H)$ and $M(T)$ curves are shown in Fig.~\ref{2_MTHErCo2}, where the abrupt change on magnetization at the magnetostructural transition (e.g. \Tc\ = 34 K at $H$ = 1 T) can be observed. The data are fully consistent with those previously reported.\cite{Burzo-1972-A, Givord-1972-A, Imai-1995-A, Duc-1999-A, Giguere-1999-A, Syshchenko-2001-A, Duc-2002-A, Oliveira-2002-A}

We have measured $ac$ magnetic susceptibility as a function of temperature (\XT) and as a function of excitation frequency \w\ (\Xw) in a variety of polycrystalline ErCo\2\ samples. The measurements were performed in a commercial SQUID Quantum Design magnetometer under zero DC applied magnetic field, from 5 to 300 K, with an excitation $ac$ field of 4.5 Oe and at \w\ ranging from 0.1 to 1000 Hz. We have also measured \XT\ under applied magnetic fields up to 2500 Oe in a single crystal, courtesy of Prof. Ernst Bauer from Vienna University of Technology. The $ac$ susceptibility measurements will be presented along the paper.

\section{Experimental results and discussion}\label{sec:Experimentalresults}

\subsection{Susceptibility as a function of temperature}\label{subsec:XT}

\begin{figure}[!b]
\begin{center}
\includegraphics[width = 0.5\textwidth]{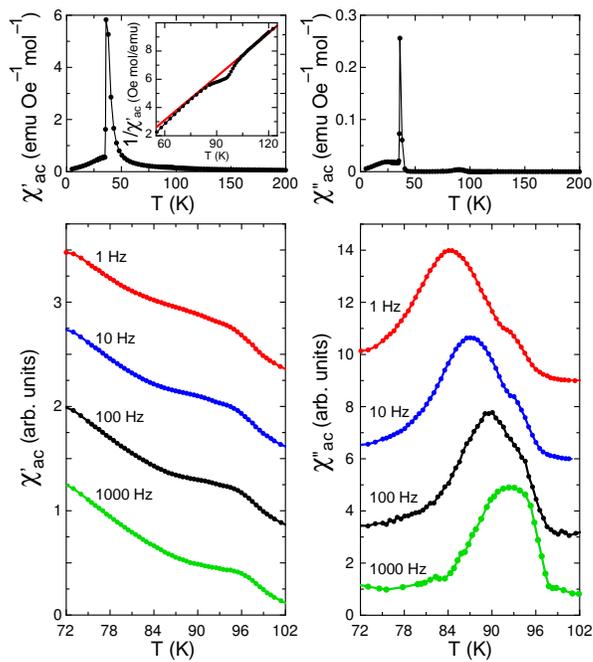}
\caption{(color online) ErCo\2\ paramagnetic $ac$ susceptibility as a function of temperature at $H$ = 0 T. Upper panels: real (\XReal) and imaginary (\XImg) components to the $ac$ susceptibility at $\omega=$ 100 Hz are shown in the left and right respectively. Lower panels:  real  and imaginary components of the $ac$ susceptibility at $\omega=$ 1, 10, 100 and 1000 Hz in the vicinity of \Tf, in this case the signals have been vertically displaced for clarity. Inset in upper left panel: inverse of \XReal\ (full dots) and its fit to a \CW\ law (continuous line) in the paramagnetic region (see text). }
\label{3_Xac_0T}
\end{center}
\end{figure}

The real (\XReal) and imaginary (\XImg) contributions to the ErCo\2\ $ac$ susceptibility as a function of temperature under zero DC applied magnetic field are shown in Fig.~\ref{3_Xac_0T}. A sharp peak can be observed at the ordering temperature (32 K) both in the real and imaginary parts, followed by an abrupt drop of the signal above \Tc. This drop corresponds to the expected \CW\ dependence of \XReal\ in the paramagnetic phase:

\begin{equation}
\chi_{ac}'(T)= \frac{C}{\mathrm{T}-\theta} \qquad \mathrm{where} \qquad C=\frac{N\mu_{B}^2}{3k_{B}} \mu_{\mathrm{eff}}^2 \label{eq:CW}
\end{equation}

However, together with that contribution to the paramagnetic $ac$ susceptibility, a small anomaly near \Tf\ = 90 K can be observed. This anomaly is more evident both in the inverse of \XReal\ and in \XImg, where a peak of 20 K-wide is present in the proximity of \Tf\ = 90 K (see inset in Fig.~\ref{3_Xac_0T}). If the region of the anomaly is excluded, the inverse of \XReal\ has a linear dependence with the temperature and therefore can be fitted to a \CW\ law using $\mu_{\mathrm{eff}}$ = 8.8 \muB, the value of the atomic Er magnetic moment. The fit, shown in the inset of upper panel in Fig.~\ref{3_Xac_0T}, is valid from high temperatures ($\sim$ 300 K) down to 30 K above the magneto structural transition, and indicates that the \CW\ contribution is due to independent Er moments. Accordingly, the difference of \XReal\ and the Er contribution ($\boldsymbol{\chi}_{Er}$) is the susceptibility due to the Co sublattice ($\boldsymbol{\chi}_{Co}$). The detachment of the different contributions to the paramagnetic susceptibility can be found in Ref.  \cite{Herrero-Albillos2007PRB76}.

The existence of a Co contribution to the $ac$ susceptibility, independent from that of Er, proves that the Co atoms have an intrinsic moment in the paramagnetic phase independent from the Er sublattice, in good agreement with the fact that temperature dependent XMCD at the Co \LL\ edges shows the presence of Co magnetic moment at temperatures above \Tc. 


Similar deviations of 1/\XReal\ from a linear dependence with the temperature well above the ordering transition has been found in other intermetallic compounds like CeNi$_{1-x}$Cu$_x$,\cite{Marcano2005PRB71, Marcano2007PRL98, Marcano2007PRB76} La$_{1-x}$Sr$_x$MnO$_3$,\cite{Deisenhofer2005PRL95} Tb$_5$Si$_2$Ge$_2$,\cite{Magen2006PRL96} Fe$_{x}$Co$_{1-x}$S$_2$ \cite{Guo2008PRL100} La$_{0.73}$Ba$_{0.27}$MnO$_3$,\cite{Jiang2007PRB76, Li2007PRB75} Nd$_{0.55}$Sr$_{0.45}$Mn$_{1-x}$Ga$_x$O$_3$,\cite{Fan2007JAP101} and Gd$_5$Ge$_4$,\cite{Ouyang2006PRB74} and have been related to the occurrence of a \GlP\ well above their ordering temperatures. The appearance of a \G\ phase is usually associated with the existence of disorder and competing interactions, which leads to percolative phenomena and clustering. Indeed SANS experiments carried in some of those compounds have demonstrated the occurrence of short range order correlations.\cite{Marcano2007PRL98, Magen2006PRL96, Yusuf2006PRB74}

In the paramagnetic phase of ErCo\2, the existence of short range order correlations is evidenced by the neutron diffraction data at 100 K presented in Fig.~\ref{1_DN} as well as in a previous SANS study on the paramagnetic phase of ErCo\2.\cite{Herrero-Albillos2007PRB76, Herrero-Albillos2007JMMM310} Those SANS measurements revealed the existence of magnetic clusters in a wide temperature range well above \Tc. 
The fits of the SANS data to lorentzian and  lorentzian-squared  functions --corresponding respectively to spin waves and static regions of spin ordering--  show that in ErCo\2\ the only 	 significant contribution is lorentzian, i.e. the short range correlations in its paramagnetic phase are due to spin waves. The correlation length, $\xi$, obtained from those fits, experiences a continuous increase from high temperatures ($>$ 200 K) down to $\approx$ 60 K (at 1 T), where it reaches a plateau at an almost constant value of 7--8 \AA\ (at lower temperatures, just above \Tc, $\xi$ diverges due to the establishment of long range order). The existence of such a plateau in $\xi$ implies that the percolation does not occur due to the growing of clusters but by the increasing of the density of clusters with sizes around 7--8 \AA. This process differs form the standard normal percolation process and has been predicted and observed for \GlP s. \cite{Bray1982JPC15, Bray1987PRL59, Salamon2003PRB68} Therefore, from what has been show above, it is reasonable to establish that both the deviation from a \CW\ law and the existence of clusters in the paramagnetic phase of ErCo\2\ comes from the presence of a \GlP.

\begin{figure}[!b]
\begin{center}
\includegraphics[width = 0.5\textwidth]{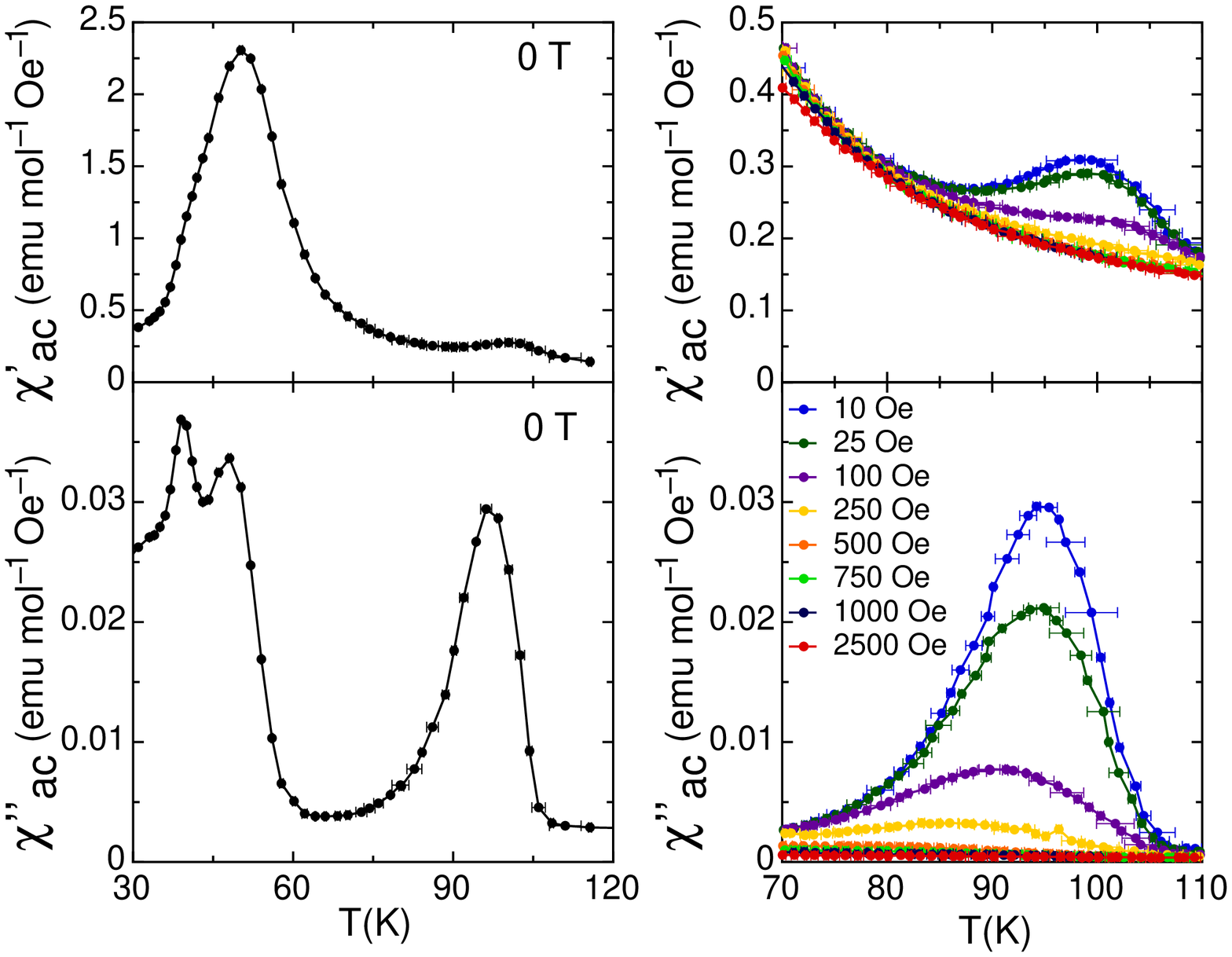}
\caption{(color online) $ac$ susceptibility as a function of temperature in an ErCo\2\ single crystal. Left graph: real (upper panel) and imaginary (lower panel) parts at $H$ = 0 T. Right graph: evolution of the real (upper panel) and imaginary (lower panel) parts in the paramagnetic region at selected applied magnetic fields up to 2500 Oe.}
\label{4_Xac_singlecrystal}
\end{center}
\end{figure}

The characteristic features of the $ac$ susceptibility in a \GlP\ are usually suppressed under small magnetic fields, an effect that has been referred to as \emph{extreme sensitivity}.\cite{Jiang2007PRB76} That is indeed the case in polycrystalline ErCo\2, where the observed peak in \XReal\ near \Tf\ = 90 K at $H$ = 0 T disappears when a magnetic field is applied. In the ErCo\2\ single crystal, the same anomaly near \Tf\ = 90 K can be also observed under moderately small applied magnetic fields. Fig.~\ref{4_Xac_singlecrystal} shows the real (upper panel) and the imaginary part (lower panel) of the susceptibility at zero applied magnetic field (left) and under applied magnetic fields up to 2500 Oe (right). A peak at \Tc\ can be observed, although it is not as sharp as in the polycrystalline samples, probably due to the twinning originated in the high pressures this sample suffered in the past, as we confirmed by x-ray Laue diffraction experiments. Nevertheless, the anomaly at \Tf\ is clearly visible both in the real and in the imaginary part, and its evolution with the applied magnetic field can be followed: the intensity of the anomaly decreases as soon as a few Oe are applied, at the same time as it moves to lower temperatures. At applied magnetic fields higher that 1000 Oe, the anomaly disappears, and the linear dependence of 1/\XReal\ with the temperature is recovered. This evolution of \XT\ with the applied magnetic field again supports the existence of a \GlP\ above the ferrimagnetic order in ErCo\2.

\subsection{Frequency dependence of the paramagnetic susceptibility} \label{subsec:Xw}

To further investigate the fingerprint of this \GlP\ in the paramagnetic susceptibility, we have also studied the dependence of the $ac$ susceptibility with the excitation frequency \w. Fig.~\ref{3_Xac_0T} shows \XT\ measurements in the vicinity of \Tf\ for \w\ = 1, 10, 100 and 1000 Hz. The anomaly already shown for 100 Hz in Fig.~\ref{3_Xac_0T} is present for all the frequencies, but a shift to higher temperatures (of 9 K when going from 1 Hz to 1000 Hz) can be observed as the frequency is increased both in the real and imaginary part or the susceptibility. This fact confirms that the origin of the anomaly at \Tf\ is not in the magnetic ordering of any impurity which may be present in the sample occurring at that temperature. Excluding the temperature region where the anomaly is present, no other frequency dependence can be observed, as the Er contribution to the paramagnetic susceptibility is \w-independent for that frequency range. The frequency dependence of the Co contribution together with the recovery of the \CW\ dependence below \Tf\ (i.e. the Co contribution going to zero below \Tf), allows to identify the origin of the anomaly as a relaxation process in the Co sublattice in the proximities of \Tf.\cite{Garcia-Palacios2006JPA38}

Therefore, the next step is to study the dynamic response of this Co contribution to the paramagnetic susceptibility. To do so, we have performed $ac$ susceptibility measurements as a function of the excitation frequency (\Xw) at zero applied magnetic field, excitation frequencies between 0.05 and 1500 Hz and in the temperature region between 80 and 100 K.

Owing to the fact that Er moments do not suffer mayor changes in that temperature range,  the Er contribution to the susceptibility can be considered as constant at fixed temperatures. This offers an advantage with respect to the study presented in the previous subsection, as it facilitates the identification of the Co contribution. i.e. Co contribution is the only contribution varying as a function of \w\ in an isothermal measurement.

\begin{figure}[!b]
\begin{center}
\includegraphics[width = 0.5\textwidth]{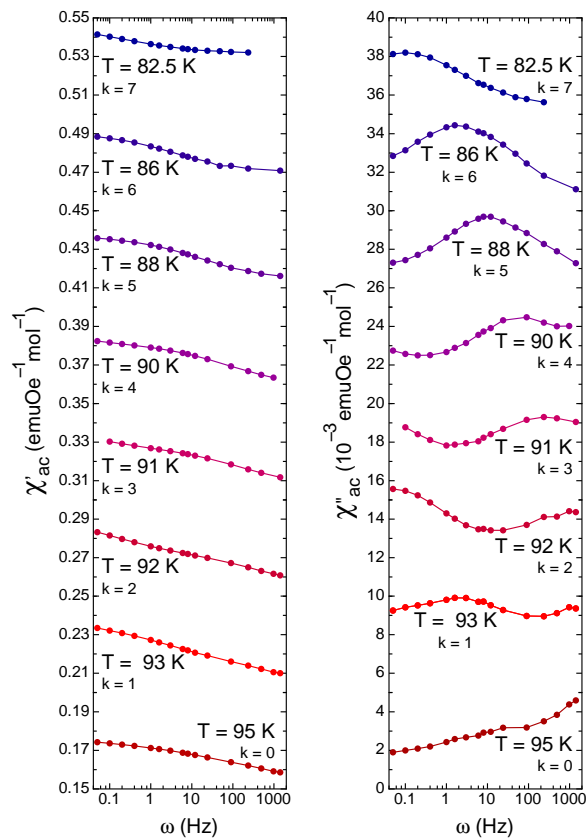}
\caption{(color online) \XReal\ (left) and \XImg\ (right) as a function of the excitation frequency at selected
temperatures between 82.5 and 95 K. The signals have been vertically displaced $k\cdot$0.05 for the real part and $k\cdot$0.005 for the imaginary part, where $k$ is indicated in each curve below the corresponding temperature.}
\label{5_Xac_frec}
\end{center}
\end{figure}

The evolution with temperature of the real and imaginary parts of \Xw\ can be observed in Fig.~\ref{5_Xac_frec}. As expected, the real part decreases as the frequency is increased. A small shoulder is present in the vicinity of \Tf, which moves to higher frequencies as the temperature is raised (in agreement with the shift observed in Fig.~\ref{3_Xac_0T}). In the imaginary contribution to \Xw, two maxima can be observed, which also move to higher frequencies as the temperature is raised.

\begin{figure}[!b]
\begin{center}
\includegraphics[width = 0.5\textwidth]{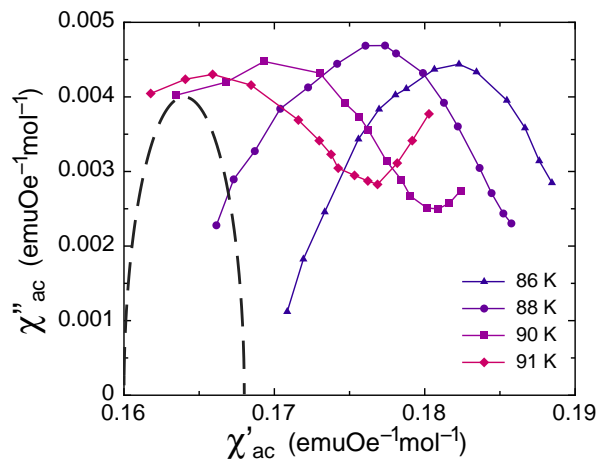}
\caption{(color online) Cole-Cole representation (i.e. \XReal\ $vs.$ \XImg) of the paramagnetic $ac$ susceptibility in ErCo\2\ at selected temperatures (full symbols). Dashed line is the curve corresponding to a theoretical Debye process with only one relaxation time.}
\label{6_ColeCole}
\end{center}
\end{figure}

Fig.~\ref{6_ColeCole} shows isothermal \XImg\ $vs.$ \XReal\ curves for selected temperatures. This kind of representation -where the excitation frequency is an implicit parameter- is the so called Cole-Cole representation. For a system of independent spins undergoing a relaxation process (i.e. a \D\ process with only one relaxation time) \XImg\ $vs.$ \XReal\ is a perfect semicircle, represented in Fig.~\ref{6_ColeCole} with a discontinuous line (note that the horizontal and vertical axes have different scales, and therefore, the circle appears to be an ellipse). This representation allows to determine, just by mere visual inspection, deviations in the susceptibility from a \D\ process: the Cole-Cole representation for ErCo\2\ shows stretched and asymmetric curves, and, in the frequency range studied, two maxima can be observed. This evidences that the observed anomaly in \Xw\ and \XT\ is not due to a single relaxation time but to a bimodal distribution of relaxation times.

The $ac$ susceptibility in a \D\ process with only one relaxation time can be usually fitted to a \D\ law\cite{Debye-1929-A}
\begin{eqnarray}\label{eq:D}
\boldsymbol{\chi}=\chi_{0}+\frac{\chi_{eq} -\chi_{0}}{1-i\omega\tau}
\end{eqnarray}
where $\tau$ is the relaxation time, $\chi_{0}$ is the value at infinite \w\ and $\chi_{eq}$ is the value at \w\ = 0, which follows a \CW\ law as a function of temperature. However, in systems with broad distributions of relaxations times as in ErCo\2, $\boldsymbol{\chi}(\omega)$ does not obey Eq.~\ref{eq:D}, and there are numerous empiric approximations to parametrize the $ac$ susceptibility. Many of those approximations are modifications of the \D\ law. In particular, the \DC\ model\cite{Davidson-1961-A, Davidson-1950-A} introduces the factor $\gamma$ ($0 < \gamma \leq 1$), which accounts for asymmetry and a smoother dependence of the susceptibility with \wt:
\begin{eqnarray}
\boldsymbol{\chi}=\chi'+i\chi''=\chi_{0}+\frac{\chi_{eq} -\chi_{0}}{(1-i\omega\tau)^\gamma}
\label{eq:DC}
\end{eqnarray}

\begin{figure}[!b]
\begin{center}
\includegraphics[width = 0.5\textwidth]{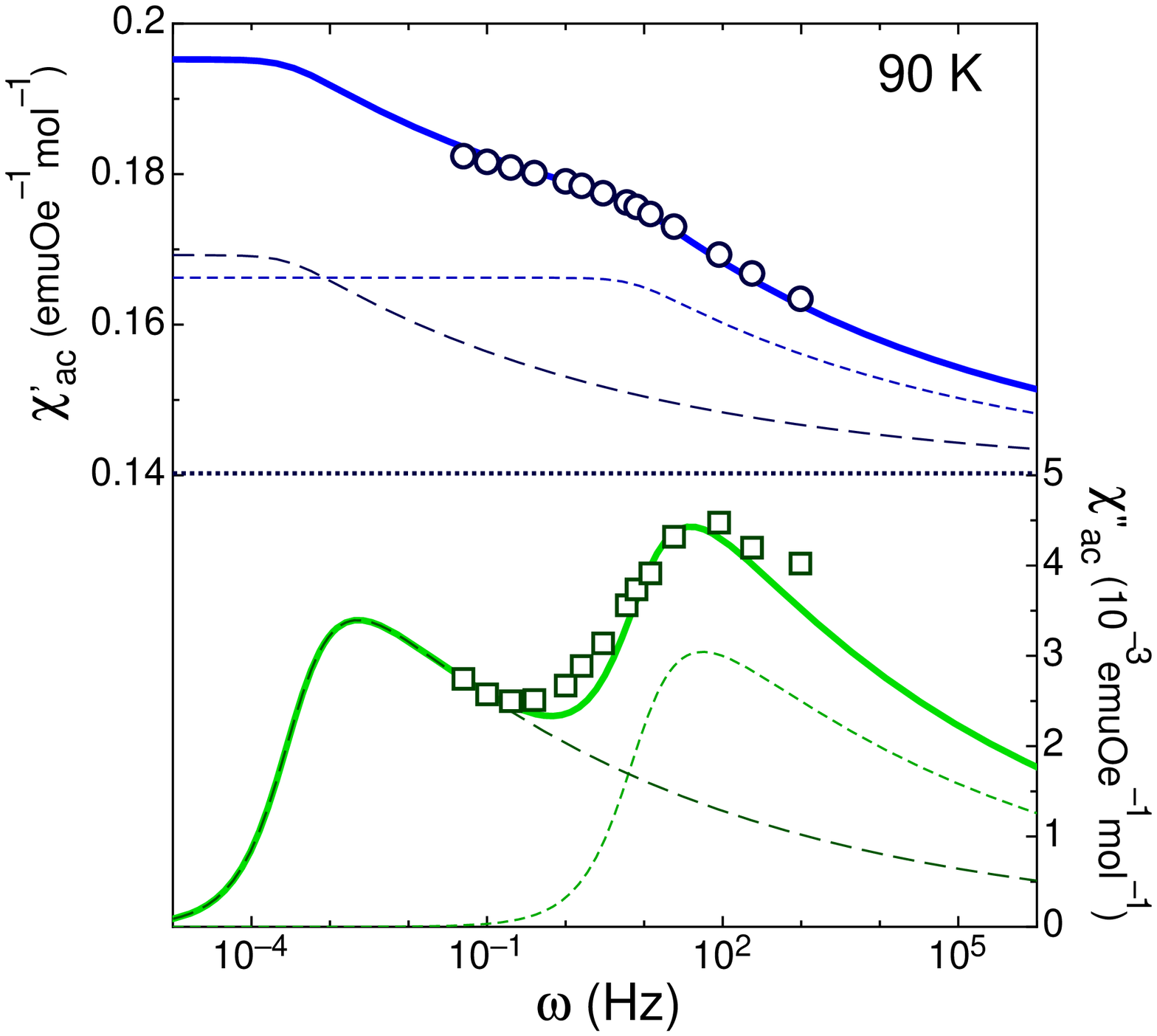}
\caption{(color online) \XReal\ (open circles) and \XImg\ (open squares) as a function of $\omega$ at 90K in ErCo\2. Continuous line is the fit of the data to a sum of two Davidson-Cole functions, the dashed lines are the two functions separately, and the dotted line is the adiabatic value of the susceptibility, $\chi_{0}$, at 90 K.}
\label{7_DavCole}
\end{center}
\end{figure}

Figure \ref{7_DavCole} shows the fit of the measured $ac$ susceptibility at 90 K to the sum of two \DC\ functions, i.e.
\begin{eqnarray}
\chi'=\chi_{0}+Re \left( \frac{\chi_{A}}{(1-i\omega\tau_{A})^\gamma} + \frac{\chi_{B}}{(1-i\omega\tau_{B})^\gamma} \right) \nonumber
\\ \nonumber
\\
\chi''=Im \left( \frac{\chi_{A}}{(1-i\omega\tau_{A})^\gamma} + \frac{\chi_{B}}{(1-i\omega\tau_{B})^\gamma} \right)
\end{eqnarray}
where $\chi_{A}$ and $\chi_{B}$ are the values of the jump in $\chi'$ for each relaxation process at $\tau_{A}$ and $\tau_{B}$. The same value of $\gamma = 0.1$ for both $\chi''$ maxima has been used, giving and idea of how broad the distribution is ($\gamma$ = 1 is the \D\ law limit).

In systems with broad distributions of relaxation times, the relaxation time frequently follows a \VF\ law.\cite{Mydosh-1993-A}
\begin{eqnarray}
\tau=\tau_{0} \exp \left(\frac{Q}{k_{B} (T-T_o)} \right) \label{eq:VF}
\end{eqnarray}
where T$_o$ is the so-called \VF\ temperature, and corresponds to the temperature at which the relaxation process occurs at zero frequency. From the \XImg\ maxima at \w\ $\tau$ = 1, the relaxation time as a function of the temperature can be obtained and from the fit of $\tau$$(T)$ to Eq.~\ref{eq:VF} T$_o$ = 70 K is obtained, which is a reasonable temperature for the zero frequency relaxation process to occur.

\subsection{Determination of the Co paramagnetic moment from susceptibility measurements}\label{subsec:calcmoment}

As mentioned earlier, $\chi_{eq}$ follows a \CW\ law with temperature and using Eq.~\ref{eq:CW} the value of the effective moment contributing the susceptibility can be calculated
\begin{eqnarray}
\mu_{\mathrm{eff}} = \left( \chi_{eq} (T-T_o) \frac{3 k_{B}}{N \mu_{B}^2 } \right) ^{1/2} \label{eq:mu}
\end{eqnarray}
where in this case N is the number of dynamic entities (i.e. number of clusters), T$_o$ is the temperature calculated form Eq.~\ref{eq:VF} and $\chi_{eq}$ can be obtained from both the fits of \Xw\ to \DC\ functions and from the \XT\ measurements. The obtained value in both cases, $\mu_{\mathrm{eff}}$~$\sim$~20~\muB, is greater than any individual moment present in the system and thus can be only explained as coming from the relaxation process of a group of moments, or cluster. Indeed, this is in good agreement with the reported cluster sizes of 7-- 8 \AA\ at \Tf\ from SANS experiments i.e. around 100 Co atoms and/or 50 Er atoms in the cluster.\cite{Herrero-Albillos2007PRB76} However, the obtained effective moment is only compatible with the SANS results if the clusters exclusively consist of Co low-spin moments. Therefore, if $c$ is the number of Co atoms in the clusters, the moment of the Co atoms in the paramagnetic phase of ErCo\2\ is $m_{\mathrm{Co}}= \mu_{\mathrm{eff}}/c$. The results are presented in table \ref{table:CoMoments}, together with the value of the Co moment obtained from an independent set of x-ray magnetic circular dichroism (XMCD) experiments presented in Ref.  \cite{Herrero-Albillos2007JMMM316}.

\begin{table}[!htb]
\begin{center}
\begin{tabular}{lr}
\hline \hline \\
Experimental data & \mCo (\muB)\\ 
\hline \\
$\chi(\omega)$ & 0.24 $\pm$ 0.03\\
$\chi(T)$ & 0.22 $\pm$ 0.03\\
XMCD\cite{Herrero-Albillos2007JMMM316} & 0.19 $\pm$ 0.02 \\
\hline \hline
\end{tabular}
\caption{Co moment in paramagnetic phase calculated from susceptibility measurements in this work and from XMCD data in Ref.  \cite{Herrero-Albillos2007JMMM316}.}
\label{table:CoMoments}
\end{center}
\end{table}

The existence of magnetic moment in the Co sublattice in the paramagnetic phase of the $R$Co\2 compounds, as well as the intrinsic or induced nature of the moment has been long studied in the literature.\cite{Gignoux-1976-A, Bloch-1970-A,Burzo-1972-A,Duc-1988-A, Liu-2006-A} The \XT\ and \Xw\ measurement presented in this work demonstrate its existence and quantifies the value of the Co magnetic moment in the paramagnetic phase of ErCo\2. What is more, the presence of an intrinsic Co component is needed to explain the relaxation process observed near \Tf. These results are also in agreement with the results from XMCD experiments\cite{Herrero-Albillos2007JMMM316} as well as with recent first principle calculations carried on ErCo\2, which demonstrate the existence of a small but yet significant Co moment at high temperatures.\cite{Herrero-Albillos2008JAP103} Moreover, the existence of a low-spin state for Co at at the onset of the ordering temperature of ErCo\2\ was also predicted theoretically by Liu and Altounian.\cite{Liu-2006-A} Our results show that the low-spin state is stable up to temperatures well above the ordering temperature \Tc.

The existence of Co intrinsic moment in the paramagnetic phase of ErCo\2\ does not agree with the calorimetric study by Imai et al.\cite{Imai-1995-A}, where the Co moment shows little influence on the magnetic entropy. However, the existence of a \GlP\ implies that the Co entropy should be released at temperatures much higher than the ordering temperature. On the other hand, the entropy as a function of temperature also discards Er moments taking part in the formation of the clusters. More detailed calorimetric studies in a larger temperature range should be performed on ErCo\2, and adequate references like ErAl\2\ and YCo\2\ should be measured to identify the influence of the formation of the Co clusters in the magnetic entropy at high temperatures.

\section{Conclusions}\label{sec:Conclusions}

$Ac$ susceptibility measurements as a function of temperature and excitation frequency have been performed in order to study the dynamic properties of the ErCo\2\ paramagnetic phase. The \XT\ measurements show the expected \CW\ dependence coming from the Er independent moments. However, a deviation form that \CW\ law, occurs in a narrow temperature region. We have studied the dependence of that anomaly with the temperature, the applied magnetic field and the excitation frequency, showing that a relaxation process occurs in the Co sublattice. The \XT\ and \Xw\ measurements also confirm that the net Co magnetic moment in ErCo\2\ paramagnetic phase is not zero and that an intrinsic component is needed to explain the relaxation process.

From the analysis of the $ac$ susceptibility measurement, together with the already reported existence of magnetic clusters in the paramagnetic phase of ErCo\2, follows that a \GlP\ is formed at temperatures well above the ordering temperature of ErCo\2, leading to a handful of new interesting phenomena in this compound. In particular, the presence of a \GlP\ is related to two new phenomena observed in ErCo\2. On one hand, the existence of a \GlP\ is related to the anomaly in the paramagnetic $ac$ susceptibility, which is due to a relaxation process of Co clusters. On the other hand, a previous work showed that the reversal of the Co magnetization occurs at a much higher temperature than the ferrimagnetic ordering and that the phenomena can be only observed if there exists Co clusters.\cite{Herrero-Albillos2007PRB76} The new phenomenology observed in ErCo\2\ is due to a competition of interactions, in which the strong Co $-$ Co exchange interaction plays a very important role, being responsible of the existence of short range order correlations in the paramagnetic phase of ErCo\2.

The generalized concept of  \GlP s  involves an arbitrary distribution of exchange constants, which can be achieved when there are competing interactions in the system toghether with a source of randomess. However, in ErCo\2, the competing interactions acting in the Co sublattice are not acompanied by a dissorder in the lattice. There are two possible sources of dissorder that could percolate the formation of the Co clusters, and therefore the formation of the \GlP.
On one hand, the appearance of small regions slightly off-stoichiometric is inherent to the synthesis process of the \RCo\ compounds (see discussion in Ref.\cite{Herrero-Albillos2006PRB73} and references therein). A slighter richer region in Co --not detectable by x-ray diffraction-- could be the seed to nucleate the clusters. Although this conjecture can not be ruled out, one would expect that, if that were the case then the phenomena would be sample dependent and  radically different in a single crystal.
On the other hand, it is well know that spin fluctuations play a very important role in the Co Laves phases. This allows us to postulate that the source of random interactions  in ErCo\2\ comes from those spin fluctuations. Indeed, the fits of the SANS measurements to a  lorentzian function rather than to a lorentzian-squared function, indicates that the Co clusters in ErCo\2\ are not  inhomogeneous regions of spin ordering but dinamical clusters.
Moreover, it is well know that the spins fluctuations are quenched when a magnetic field is applied\cite{Ikeda1984PBR29, Ikeda1991JMMM100}, in agreement with the already mentioned extreme sensibility of the  \GlP\ to the applied magnetic field. 

However, although the occurrence of a \GlP\ in ErCo\2\ has some distinct characteristic, the scenario is very similar to what it has been reported in other intermetallic compounds. Indeed, there is an increasing amount of compounds in which the characteristic fingerprint of \GlP s are now being reported as manganites, rare-earth alloys, heavy-fermion systems, itinerant magnetic semiconductors and now the case of ErCo\2. It is therefore very likely that if the paramagnetic phase of other well studied intermetallic compounds are revisited, several of them would show \G\ like behavior. The characterization of those compounds would surely help in the understanding of both the magnetic properties of those particular systems and of the paramagnetic phase of intermetallic compounds in general.

\section*{Acknowledgments}
We acknowledge the Spanish CICYT research project MAT2005-02454, the FEDER program and the Aragon\'{e}se DECRYPT and CAMRADS research groups. We thank F. Luis for fruitful discussion, N. Plugaru and M. J. Pastor for sample preparation and Prof. E. Bauer for the single crystal. J.H. acknowledge Fundaci\'{o}n Ram\'{o}n Areces for financial support.


\end{document}